\documentclass[twocolumn,tightenlines,aps,preprintnumbers,superscriptaddress,showpacs,floatfix,nofootinbib]{revtex4}

\usepackage{amssymb}
\usepackage{amsmath}
\usepackage{epsfig}

\usepackage{epsf}
\usepackage{graphicx}
\usepackage{amsfonts}
\usepackage{amsthm}
\usepackage{mathrsfs}
\usepackage{wrapfig}
\usepackage{color}

\pdfoutput=1

\def\w{{\omega}}
\def\a{{\alpha}}
\def\b{{\beta}}
\def\e{{\epsilon}}

\newcommand{\pd}{\partial}
\newcommand{\tr}{\text{tr}\,}

\newcommand{\la}{\langle}
\newcommand{\ra}{\rangle}

\newcommand{\diag}{\text{diag}}

\newcommand{\be}{\begin{equation}}
\newcommand{\ee}{\end{equation}}
\newcommand{\bea}{\begin{eqnarray}}
\newcommand{\eea}{\end{eqnarray}}
\newcommand{\barr}{\begin{array}}
\newcommand{\earr}{\end{array}}

\newcommand{\bigO}[1]{{\mathcal O\left(#1 \right)}}

\def\be{\begin{equation}}
\def\ee{\end{equation}}
\def\bea{\begin{eqnarray}}
\def\eea{\end{eqnarray}}

\def\ra{\rangle}
\def\la{\langle}

\begin{document}

\title{Entanglement Entropy of Periodic Sublattices}
\author{Temple He}
\affiliation{Center for the Fundamental Laws of Nature, Harvard University, Cambridge, MA 02138, USA}
\author{Javier M. Mag\'an}
\affiliation{Institute for Theoretical Physics and Center for Extreme Matter and Emergent Phenomena, Utrecht University, 3508 TD Utrecht, The Netherlands}
\author{Stefan Vandoren}
\affiliation{Institute for Theoretical Physics and Center for Extreme Matter and Emergent Phenomena, Utrecht University, 3508 TD Utrecht, The Netherlands}

\date{\today}

\begin{abstract}

We study the entanglement entropy (EE) of Gaussian systems on a lattice with periodic boundary conditions, both in the vacuum and at nonzero temperatures. By restricting the reduced subsystem to periodic sublattices, we can compute the entanglement spectrum and EE exactly. We illustrate this for a free (1+1)-dimensional massive scalar field at a fixed temperature. Consistent with previous literature, we demonstrate that for a sufficiently large periodic sublattice the EE grows extensively, even in the vacuum. Furthermore, the analytic expression for the EE allows us probe its behavior both in the massless limit and in the continuum limit at any temperature.

\end{abstract}


\maketitle

\section{Introduction}

Entanglement entropy (EE) of Gaussian systems has been studied thoroughly in the literature \cite{Bombelli:1986rw,srednickisub,Holzhey:1994we,Plenio:2004he,Calabrese:2004eu,Casini:2005zv,Eisert:2008ur,Casini:2009sr,Calabrese:2009qy,Peschel_Eisler,Herzog:2012bw}, and the focus of this paper is mostly on one spatial dimension. Much of the literature involves calculating the EE when the reduced subsystem $A$ is taken to be one or more line intervals or, in the discretized case, one or more sets of adjacent lattice points. Except for a few cases, such as a conformal field theory \cite{Holzhey:1994we,Calabrese:2004eu,Calabrese:2009qy,Cardy:2014jwa} or a massive scalar on $\mathbb{R}^2$ at zero temperature \cite{Calabrese:2004eu,Casini:2005zv,Casini:2009sr}, the EE is notoriously hard to compute analytically, and one often relies on numerical approaches.

The question naturally arises whether the EE can be computed exactly if we did not restrict our subsystem $A$ to a line interval or adjacent lattice points. In this paper, we examine this question by studying Gaussian systems on a lattice with periodic boundary conditions. They can arise from field theories compactified on a spatial circle of circumference $L$, in which we regulate the UV divergence by discretizing the circle into $N$ lattice sites with lattice separation $\e = L/N$. Concretely, we will consider a free (1+1)-dimensional massive scalar field discretized on a lattice, but the general idea can be applied to any Gaussian quantum many-body system with translation symmetry. 

\begin{figure}[h]
\centering
{\includegraphics[height=30mm]{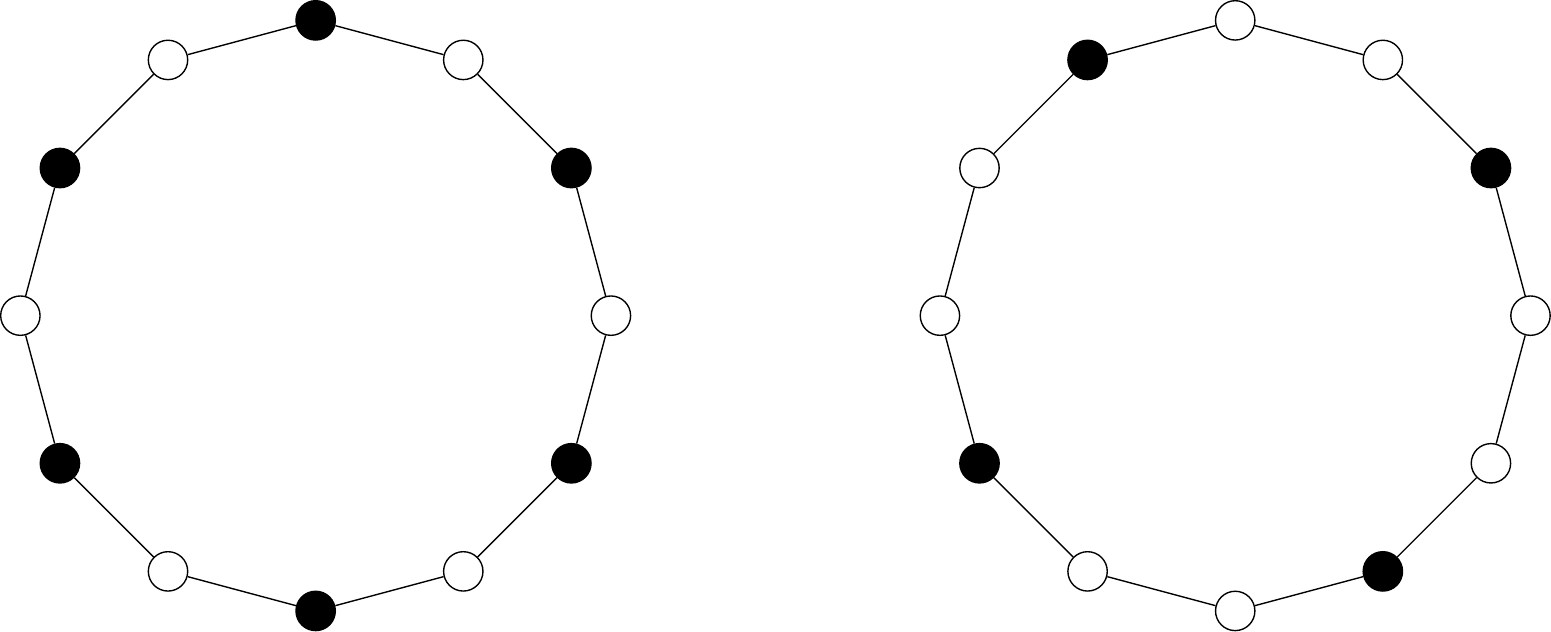}}

\caption{{\small We depict as an example a one-dimensional periodic lattice with $N=12$, with subsystem $A$ consisting of the filled lattice points. On the left $N_A=6$ and $p=2$, while on the right $N_A=4$ and $p=3$.}}
\label{fig1}
\end{figure}

To allow an analytical treatment, we choose the subsystem $A$ to be a periodic sublattice, consisting of $N_A$ evenly spaced lattice points such that 
\begin{align}
	N_A=\frac{1}{p}N, \quad  p \in \mathbb Z^+\ .
\end{align}
This is summarized in Figure \ref{fig1}.

The advantage of choosing our subsystem $A$ in this way is that discrete translation symmetry along the circle is preserved. This will allow us to compute the EE $S_{A}$ exactly, despite the fact that it can be seen (or used) as a notion of multipartite entanglement. Exact solvability is rooted in two features.  Firstly, for Gaussian systems the EE can be obtained solely from the two-point correlation functions between the lattice sites of $A$ \cite{Peschel1,Vidaletal}. Secondly, for such periodic sublattices we show that the two-point correlation matrix can be diagonalized using theorems about eigenvalues of circulant matrices.

While part of our motivation is to simply have new quantum systems whose EE can be computed exactly, sublattice EE is an interesting field of study in its own right. For instance, sublattice EE with $p=2$ in spin chains was shown to be an order parameter that could signal quantum phase transitions \cite{sublattice-entanglement,sublattice-entanglement2,IP}. Furthermore, in one spatial dimension, sublattice EE grows extensively with the size of the sublattice even in the vacuum, in contrast to the usual logarithmic growth \cite{IP}.\footnote{For small subsystems, there might be deviations from extensivity. An example is the transverse Ising model on a ring in the ordered region \cite{IP}. Deviations from extensivity will also arise in our setup, but only for systems with few lattice sites.}  We confirm this in our bosonic models as well, at least for sublattices $A$ with sufficiently many lattice sites. 
Extensivity arises because the number of interface points between $A$ and its complement is equal to the size of $A$. This continues to hold in the continuum limit, and it would be interesting to understand this holographically.

\begin{figure}[h]
\centering
{\includegraphics[height=30mm]{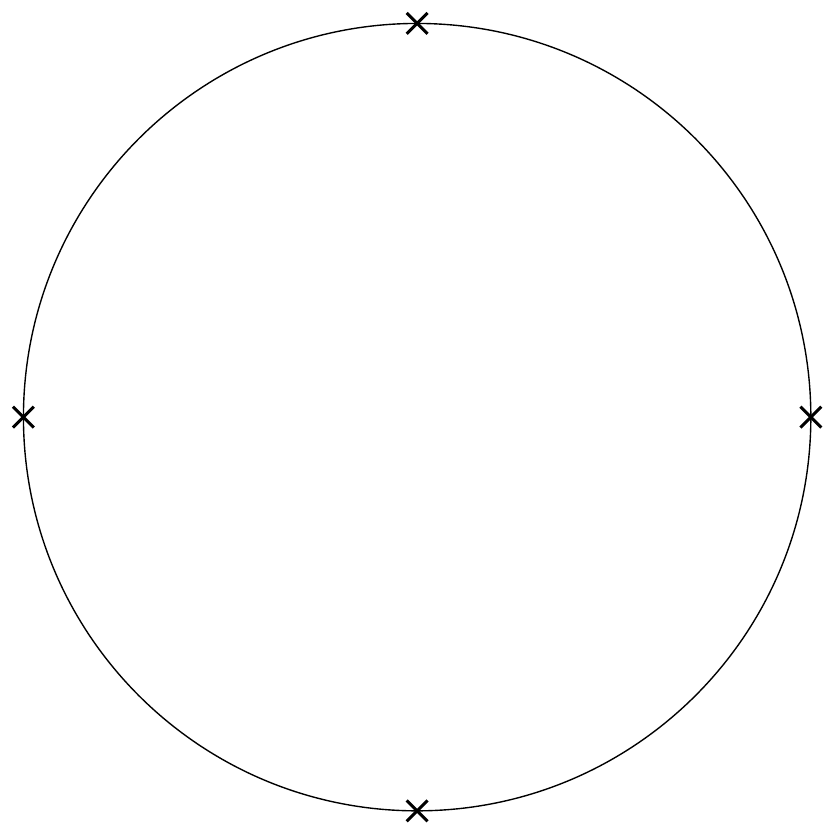}}

\caption{{\small We illustrate the continuum limit where $p,N\rightarrow \infty$ while keeping $N/p=N_A$ fixed. In the figure, $N_A=4$.}}
\label{fig2}
\end{figure}

Finally, we are also motivated to study the behavior of sublattice EE in various limits, such as the massless limit, the continuum limit, and the high and low temperature limits. The continuum limit we will consider here is one where both $N$ and $p$ are taken to infinity with $N_A$ fixed. This setup entangles a sparse set of points with the rest of the degrees of freedom living on the circle, as shown in Figure \ref{fig2}.

The organization of this paper is as follows. In Section 2, we present our model and discuss some relevant properties of circulant matrices. In Section 3, we review the general techniques to compute the EE exactly for Gaussian systems. These techniques are applied to periodic sublattices in Section 4, which forms the bulk of this paper and is where the novel results are presented.

\section{Hamiltonian, Correlators and Circulant Matrices}

While the techniques developed below are generally applicable to arbitrary Gaussian systems,\footnote{See Section \ref{discussion} for a more detailed explanation of this point.} we choose for concreteness to study a (1+1)-dimensional massive scalar field with Hamiltonian
\begin{equation}
H = \frac{1}{2}\int_0^L dx\,\left[\pi(x)^2 + (\pd_x\phi(x))^2+m^2\phi(x)^2\right]\ .
\end{equation}
The space is compact with length $L$ and we impose periodic boundary conditions $\phi(0)=\phi(L)$. We then discretize the theory by letting $x_j = j\e$ be the allowed positions on the circle, with $j = 0,1,\ldots,N-1$ and $L=N\e$. Denoting $\phi(j\e) \equiv \phi_j$ and $\pi(j\e) \equiv \pi_j/\e$ so that both $\phi_j$ and $\pi_j$ are dimensionless, the Hamiltonian becomes
\begin{equation}\label{discrete-Ham}
 	H = \frac{1}{2\e}\sum_{j=0}^{N-1} \left[\pi_j^2 + (\phi_{j+1} - \phi_j)^2+m^2\epsilon^2\phi_j^2\right]
\end{equation}
with periodic boundary condition $\phi_{0}=\phi_N$. This is just a set of coupled harmonic oscillators, which has been considered already in \cite{Bombelli:1986rw,srednickisub} and in many subsequent papers. The ground state wave function is normalizable and well-defined only if $m\neq 0$. The dimensionless parameters in this theory are $m\e$, $N$, and $\beta m$, where $\beta$ is the inverse temperature. The reduced density matrix also depends on $N_A$, so we have four dimensionless parameters that the EE depends upon.

As we will review in detail in the next section, the calculation of the EE for Gaussian systems can be computed from the two-point correlation functions of fields evaluated within the subsystem. For our model, the relevant correlators between any two points on the lattice are well known from the literature,\footnote{They can also be derived directly for free theories by expanding $\phi_i$ into creation and annihilation operators.} and we follow the notation used in \cite{Herzog:2012bw}. We will in particular only be interested in the correlators between points on the sublattice that are labeled by $0, p,\ldots,(N_A-1)p$. By rescaling the indices by $p$ so that they instead run from $0$ to $N_A-1$, the correlators on the sublattice then take the form
\begin{align}
\begin{split}\label{corr}
	\Phi_{ij}\equiv \la \phi_i\phi_j \ra&= \frac{1}{2N}\sum_{k=0}^{N-1} \frac{1}{\e\omega_k}(2 \la N_k\ra +1)\cos \frac{2\pi(i-j)k}{N_A}\ , \\
	\Pi_{ij}\equiv \la \pi_i\pi_j \ra &= \frac{1}{2N}\sum_{k=0}^{N-1} \e\omega_k(2\la N_k \ra+1)\cos  \frac{2\pi(i-j)k}{N_A}\ ,
\end{split}
\end{align}
with $i,j=0,1,\ldots,N_A-1$, and $\w_k$ is given via the dispersion relation
\begin{align}\label{disp}
	\w_k^2 = m^2 + \frac{4}{\e^2}\sin^2\frac{\pi k}{N}\ .
\end{align}

The correlators in \eqref{corr} are true in any state for which $\la a_i a_j \ra = \la a^\dagger_i a^\dagger_j \ra=0$, where $a_i$ and $a^\dagger_j $ are the annihilation and creation operators appearing in the mode expansion of the scalar field. These conditions are for instance satisfied in the vacuum state and in the thermal state with inverse temperature $\b$, the latter following from the fact
\begin{align}
	\tr[\rho_\beta \,a_i\, a_j]= \tr[\rho_\beta \,a_i^\dagger\, a_j^\dagger]=0\ ,
\end{align}
where $\rho_\beta$ is the thermal density matrix. They also are satisfied in typical states subjected to random unitary dynamics, see \cite{vijay,MV}. The average number of particles with momentum mode $k$ in the thermal state is given by the Bose-Einstein distribution:
\begin{equation}\label{Bos-Ein}
\la N_k \ra =\frac{1}{e^{\beta \omega_k}-1}\ .
\end{equation}

Observe that the correlation matrices in \eqref{corr} only depend on the difference $i-j$; such matrices are called Toeplitz matrices. In fact, since we are considering periodic sublattices, the correlation matrices fall into an even more restrictive class of matrices known as \emph{circulant matrices}. A circulant matrix is a matrix where every row is a cyclic right shift of the row above it, so that the entire matrix is determined by its first row. For instance, the $3\times 3$ matrix
\begin{equation}
	C=\begin{pmatrix} c_0 & c_1 & c_2 \\ c_2 & c_0 & c_1 \\ c_1 & c_2 & c_0 \end{pmatrix}
\end{equation}
is circulant, and is commonly denoted in the literature by its first row as $C=\text{circ}(c_0,c_1,c_2)$.

Circulant matrices have many desirable properties. For instance, the sum or the product of two circulant matrices is again circulant. Furthermore, their eigenvalues and eigenvectors can be determined rather easily; for an $N_A\times N_A$ circulant matrix $\text{circ}(c_0,c_1,\ldots,c_{N_A-1})$ the eigenvalues are given by
\begin{equation}\label{circ-eigenvalue}
	\lambda_m=\sum_{k=0}^{N_A-1}c_k\,e^{-2\pi im k/N_A}\ , \quad m=0,1,\ldots,N_A-1\ ,
\end{equation}
and the corresponding eigenvectors are 
\begin{equation}\label{circ-eigenvector}
	v^{(m)}=\frac{1}{\sqrt {N_A}}\left(1,e^{-2\pi i m/N_A},\cdots, e^{-2\pi i m (N_A-1)/N_A}\right)^T\ .
\end{equation}
Notice that circulant matrices all have the same eigenvectors, so that circulant matrices commute and can be diagonalized simultaneously. This is the key feature that will allow the analytical treatment of EEs, as we show in the next section. For some references on circulant matrices, see \cite{PJDavis,RMGray}.

We conclude this section by pointing out that the reason the correlation matrices in our model are circulant is due to the structure of our Hamiltonian. Our model of a discretized massive scalar field theory on a circular lattice belongs to a more general class of coupled harmonic oscillators with Hamiltonian
\begin{equation}\label{discrete-Ham-circ}
 	H = \frac{1}{2\e}\left(\sum_{j=0}^{N-1} \pi_j^2 +\sum_{i,\,j=0}^{N-1} \phi_iV_{ij}\phi_j\right)\ ,
\end{equation}
where $V_{ij}$ is a symmetric coupling matrix. Translational invariance then implies that $V$ is Toeplitz, and the periodic boundary condition $\phi_{0}=\phi_N$ implies that $V$ is circulant. The Hamiltonian \eqref{discrete-Ham} is a special case of \eqref{discrete-Ham-circ} with nearest-neighbor interactions
\begin{equation}\label{V-KG}
	V={\text{circ}}\left(2+ m^2\e^2,-1,0,\ldots,0,-1\right)\ .
\end{equation}
The two-point correlation matrices of a theory are circulant if $V$ is circulant. However, for general subsystems, the correlation matrices restricted to the reduced subsystem will only be Toeplitz. The general idea behind our method is to choose a subsystem that again respects periodicity, so that the correlation matrices restricted to the subsystem are still circulant. It is easy to check that the correlation matrices in \eqref{corr} are circulant as, for a fixed integer $k$, the $N_A\times N_A$ matrix $A_{mn}^{(k)}=\cos\left(\frac{2\pi k}{N}(m-n)\right)$ is circulant. This idea has a broad range of applicability, since the interactions need not be restricted to nearest-neighbor interactions as in \eqref{V-KG}, but can also include long-range interactions as long as $V$ is circulant.

\section{Entanglement Spectrum and the Modular Hamiltonian}
For Gaussian systems, the EE can be computed from the two-point correlators \cite{Peschel1, Vidaletal,Botero}. We will use this so-called real time approach to compute the EE instead of the Euclidean approach where the popular replica trick is often utilized, though both approaches are directly related \cite{Casini:2009sr}. Although we will be mainly focused on bosonic systems, many of the concepts presented in this section can be straightforwardly extended to fermionic Gaussian systems as well.

We begin by constructing the vector consisting of positions and momenta living on the sublattice
\begin{align}
 	\eta_\alpha \equiv (\phi_0,\ldots,\phi_{N_A-1}, \pi_0,\ldots,\pi_{N_A-1})\ ,
\end{align}	
where $\alpha=1,\ldots,2N_A$. This allows us to introduce a $2N_A \times 2N_A$ covariance matrix $C$ with components \cite{SMD}
\begin{equation}\label{Cab}
 	C_{\a\b} \equiv \frac{1}{2}\la\{\eta_\a,\eta_\b\} \ra = \frac{1}{2}\begin{pmatrix}
 		\la\{\phi_i,\phi_j\}\ra & \la\{\phi_i,\pi_j\}\ra \\
 	\la\{\pi_i,\phi_j\}\ra & \la\{\pi_i,\pi_j\}\ra
 	\end{pmatrix}\ ,
\end{equation}
where $i,j$ run from $0,\ldots,N_A-1$ and $\{\cdot,\cdot\}$ is the anticommutator. As this is an even-dimensional positive-definite matrix, Williamson's theorem \cite{Williamson} states that there exists a symplectic transformation $S_W$ that diagonalizes $C$, i.e.
\begin{align}\label{w}
 	W \equiv S_W^T C S_W = \diag(\lambda_0,\ldots,\lambda_{N_A-1},\lambda_0,\ldots,\lambda_{N_A-1})\ .
\end{align}
Here the $\lambda_l$'s are \emph{not} the eigenvalues of $C$, since eigenvalues are not preserved under a symplectic transformation. They are instead what are known as symplectic eigenvalues. These symplectic eigenvalues of $C$ are doubly degenerate, and it can be shown that they are all at least $\frac{1}{2}$ as a consequence of the uncertainty principle \cite{SMD}.

The reason these symplectic eigenvalues are interesting is because they allow us to determine the reduced density matrix \cite{Peschel1,Botero}. Under the symplectic transformation, $\eta$ is mapped to $\tilde\eta \equiv S_W\eta$. Since $W$ is diagonal, we can decompose the components of $\tilde\eta$ into creation and annihilation operators:
\begin{align}\label{modes}
\ 	\tilde\phi_i &= \frac{\tilde a^\dagger_i + \tilde a_i}{\sqrt 2}\ , \qquad \tilde\pi_i = i\frac{\tilde a^\dagger_i - \tilde a_i}{\sqrt 2}\ .
\end{align}
In this basis, the reduced density matrix factorizes as \cite{Peschel1,Vidaletal}
\begin{align}\label{density_matrix}
 	\rho_A = \bigotimes_{l=0}^{N_A-1}\rho_l\ ,\qquad \rho_l= (1-e^{-\beta_l})e^{-\beta_l\tilde n_l}\ ,
\end{align}
where $\tilde n_l = \tilde a_l^\dagger \tilde a_l$ is the Williamson number operator. From the parameters $\beta_l$ the modular Hamiltonian $H_A$, defined via the equation
\begin{align}
	\rho_A= \frac{e^{-H_A}}{\tr \left(e^{-H_A}\right)} \ ,
\end{align} 
can be written as
\begin{equation}\label{mod_ham}
	H_A=\sum_{l=0}^{N_A-1}H_l\ ,\qquad H_l=\beta_l\,\tilde a_l^\dagger \,\tilde a_l\ .
\end{equation}
Furthermore, one can establish that $\b_l$ is related to the symplectic eigenvalue $\lambda_l$ via the relation
\begin{align}\label{b_i2}
	\b_l = \log\frac{\lambda_l + \frac{1}{2}}{\lambda_l - \frac{1}{2}} \quad\Leftrightarrow\quad \lambda_l=\frac{1}{2}\coth\left(\frac{\beta_l}{2}\right)\ .
\end{align}
Because the eigenvalues of $\rho_l$ are
\begin{align}\label{probs}
 	p_n^{(l)} = \left(1-e^{-\b_l}\right)e^{-\b_ln}, \quad n=0,1,\ldots\ ,
\end{align}
and the EE is by definition
\begin{align}
	S_A = -\sum_{l=0}^{N_A-1} \tr\left(\rho_l\log \rho_l\right) = -\sum_{l=0}^{N_A-1}\sum_{n=0}^\infty p_n^{(l)}\log p_n^{(l)}\ ,
\end{align}
applying \eqref{b_i2} and \eqref{probs} leads directly to the equation
\begin{align}\label{S_herzog}
\begin{split}
 	S_A &= \sum_{l=0}^{N_A-1} \left[\left(\lambda_l + \frac{1}{2} \right)\log\left(\lambda_l + \frac{1}{2} \right) \right. \\
 	&~~~~~~~~~~~~~~~~~~~~ \left. - \left(\lambda_l - \frac{1}{2} \right)\log\left(\lambda_l - \frac{1}{2} \right)\right]\ .
\end{split}
\end{align}
Note that in order for this equation to make sense, we indeed require that the symplectic eigenvalues are greater than or equal to $\frac{1}{2}$. We also remark that for large $\lambda_l$, the term in the summation simplifies to
\begin{equation}\label{Slargelambda}
	\log\lambda_l + 1 + \bigO{1/\lambda_l^2}\ .
\end{equation}

We remark here that some of the literature, i.e. \cite{Bombelli:1986rw,srednickisub}, also express the EE in terms of the parameters
\begin{equation}\label{xi_param}
	\xi_l \equiv \frac{\lambda_l-\frac{1}{2}}{\lambda_l+\frac{1}{2}} = e^{-\b_l}\ .
\end{equation}
When $\lambda_l \rightarrow \frac{1}{2}$, we have $\xi_l \rightarrow 0$ and when $\lambda_l\rightarrow \infty$ we have $\xi_l \rightarrow 1$. In terms of these variables we can write the EE as 
\begin{equation}\label{S_xi}
	S_A=\sum_{l=0}^{N_A-1}S_A^{(l)}\ , \quad S_A^{(l)} \equiv -\log(1-\xi_l)-\frac{\xi_l}{1-\xi_l}\log\xi_l\ .
\end{equation}
The EE diverges when $\xi_l\rightarrow 1$ for some $l$. This happens in the massless limit, as well as in the infinite temperature limit.

As an intermezzo, it is very instructive to prove the first law of EE using the symplectic eigenvalues, that is,
\begin{align}\label{EE_first}
	\delta S_A = \tr\left( \delta\rho_A H_A \right) = \delta \la H_A \ra  \ ,
\end{align}
where $H_A$ is the modular Hamiltonian. First, we compute $\delta S_A$ using \eqref{S_herzog} and obtain
\begin{equation}\label{var-See}
	\delta S_A=\sum_{l=0}^{N_A-1}\beta_l\, \delta \lambda_l\ .
\end{equation}
Next, from \eqref{density_matrix} and \eqref{mod_ham} we get
\begin{equation}\label{tr_rhoH}
	\tr(\delta \rho_A H_A)=\sum_{l=0}^{N_A-1}\sum_{n=0}^{\infty}\delta p_{n}^{(l)}\, \beta_l\,  n \ .
\end{equation}
Using the relations \eqref{b_i2} and \eqref{probs} and after some algebra, we again arrive at the right-hand side of \eqref{var-See}, thus proving the first law of EE. To our knowledge, \eqref{var-See} has not appeared in the literature before, and it provides a novel and efficient way of computing variations in the EE. In the next section, we will exploit it to compute the variation of EE under an infinitesimal temperature deviation from the vacuum.

Thus far, everything in this section applies to any generic Gaussian systems. For the Hamiltonian \eqref{discrete-Ham} however, the covariance matrix further simplifies because there are no correlations between $\phi_i$ and $\pi_j$, as can be checked using mode expansions. Therefore, $C_{\a\b}$ is in fact block diagonal:
\begin{align}\label{v}
\begin{split}
 	C_{\a\b} &=\frac{1}{2}\begin{pmatrix}
 		\la\{\phi_i,\phi_j\}\ra &0 \\
 	0 & \la\{\pi_i,\pi_j\}\ra
 	\end{pmatrix}  \\
 	&= \begin{pmatrix}
 		\la\phi_i\phi_j\ra &0 \\
 	0 & \la\pi_i\pi_j\ra
 	\end{pmatrix} \\
 	&= \begin{pmatrix}
 		\Phi_{ij} &0 \\
 	0 & \Pi_{ij}
 	\end{pmatrix}\ ,
\end{split}
\end{align}
where $\Phi_{ij}$ and $\Pi_{ij}$ were defined in \eqref{corr}. We used the fact $\phi_i$ and $\phi_j$ commute for any $i$ and $j$ in the second equality, so $\{\phi_i,\phi_j\} = 2\phi_i\phi_j$; the analogous statement holds for the $\pi_i$'s.

In general, given a Toeplitz matrix, it is not always easy to determine its symplectic eigenvalues. However, because $C$ is block diagonal, it has been shown that its symplectic eigenvalues (ignoring the double degeneracy) are precisely the eigenvalues of the matrix $\sqrt{\Phi\Pi}$ \cite{Peschel1,Botero}. 

Our task is further simplified by the fact that the matrices $\Phi$ and $\Pi$ are circulant and are hence simultaneously diagonalizable. Thus, if $\lambda^\phi_l$ are the eigenvalues of $\Phi$ and $\lambda^\pi_l$ are those of $\Pi$, the eigenvalues of $\sqrt{\Phi\Pi}$, and hence the symplectic eigenvalues of $C$, are given by 
\begin{align}\label{sym_eval}
	\lambda_l = \sqrt{\lambda^\phi_l\lambda^\pi_l}\ .
\end{align}
As the eigenvalues of circulant matrices are given in \eqref{circ-eigenvalue}, we can now determine the $\lambda_l$'s via \eqref{sym_eval}, which will in turn let us determine the EE via \eqref{S_herzog}. We present the results in the next section.

\section{Results}

As we mentioned in the previous section, the fact $\Phi$ and $\Pi$ are circulant matrices implies that their eigenvectors are both given by \eqref{circ-eigenvector}. We can thus compute their respective eigenvalues $\lambda^\phi_l$ and $\lambda^\pi_l$ directly from \eqref{corr} using the general formula for the eigenvalues given in \eqref{circ-eigenvalue}. After some algebra, we obtain (recall that $p = N/N_A$) 
\begin{widetext}
\begin{align}\label{lambda}
\begin{split}
	\lambda^\phi_l &= \frac{1}{4p}\sum_{k=0}^{p-1} \left[\frac{1}{\e\w_{l+kN_A}}(2\la N_{l+kN_A}\ra+1)  + \frac{1}{\e\w_{(k+1)N_A-l}}(2\la N_{(k+1)N_A-l} \ra + 1) \right] \ ,\\
	\lambda^\pi_l &= \frac{1}{4p}\sum_{k=0}^{p-1} \left[\e\w_{l+kN_A}(2\la N_{l+kN_A}\ra+1)  + \e\w_{(k+1)N_A-l}(2\la N_{(k+1)N_A-l} \ra + 1) \right]\ .
\end{split}
\end{align}
\end{widetext}
The symplectic eigenvalues are thus for $l = 0,\ldots,N_A-1$
\begin{widetext}
\begin{align}\label{ex_sym}
\begin{split}
	\lambda_l &= \frac{1}{4p} \left[\sum_{j=0}^{p-1} \left( \frac{2\la N_{l+jN_A} \ra +1}{\w_{l+jN_A}} + \frac{2\la N_{(j+1)N_A-l}\ra+1}{\w_{(j+1)N_A-l}}   \right) \right. \\
 	&~~~~~~~~~ \times \left.\sum_{k=0}^{p-1}\left(  (2\la N_{l+kN_A}\ra + 1)  \w_{l+kN_A} + (2\la N_{(k+1)N_A-l}\ra + 1)  \w_{l+kN_A}\right)  \right]^{1/2}\ ,
\end{split}
\end{align}
\end{widetext}
and substituting them into \eqref{S_herzog} yields the EE of subsystem $A$. 

If we assume that the system is in a thermal state with inverse temperature $\b$, then we can use \eqref{Bos-Ein} to write the symplectic eigenvalues as
\begin{widetext}
\begin{align}\label{therm_lambda}
\begin{split}
	\lambda_l &= \frac{1}{4p} \left[\sum_{j=0}^{p-1} \left( \frac{\coth\frac{\b\w_{l+jN_A}}{2}}{\w_{l+jN_A}} + \frac{\coth\frac{\b\w_{(j+1)N_A-l}}{2}}{\w_{(j+1)N_A-l}}   \right) \right. \\
 	&~~~~~~~~~ \times \left.\sum_{k=0}^{p-1}\left(\w_{l+kN_A}  \coth\frac{\b\w_{l+kN_A}}{2}   + \w_{(k+1)N_A-l} \coth\frac{\b\w_{(k+1)N_A-l}}{2} \right)  \right]^{1/2}\ .
\end{split}
\end{align}
\end{widetext}
The result for the vacuum state is much simpler, and follows from taking the zero temperature limit $\beta\rightarrow\infty$ in \eqref{therm_lambda}:
\begin{align}\label{mass_lambda}
	\lambda_l = \frac{1}{2p} \left[\,\,\sum_{j,k=0}^{p-1}\frac{\w_{l+kN_A}}{\w_{l+jN_A}}\right]^{1/2}\ ,\qquad l=0,1,\ldots, N_A-1\ .
\end{align}
One special case to consider is for $N_A=1$ (so $p=N$) in the vacuum, i.e. we are entangling a single lattice site with the rest of the system, and it was studied in \cite{Mallayya:2014xwa}. Here we extend it to general values of $p$ and nonzero temperatures. To gain more intuition regarding this result, we will focus on several limiting cases. First, we look at a system where $p=2$, i.e. our subsystem consists of every other point on the circular lattice. We then turn to the opposite limit when $p \gg 1$.

\subsection{Two Coupled Harmonic Oscillators}

\begin{figure}[h]
\centering
{\includegraphics[height=30mm]{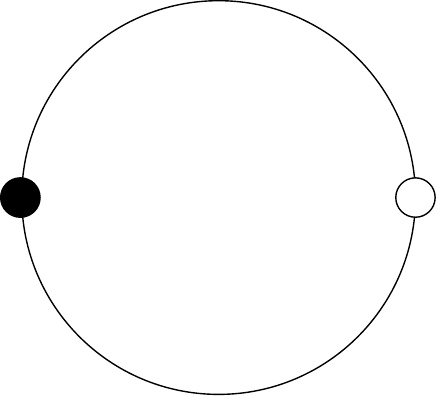}}

\caption{{\small Two coupled harmonic oscillators, corresponding to $N=2, N_A=1$.}}
\label{fig3}
\end{figure}

As a warm-up, we begin our analysis of the case when $p=2$ with only two points on the circle. Then our subsystem $A$ consists of only one point, as depicted in Figure \ref{fig3}.

In this simple case, there is only one symplectic eigenvalue, which in the thermal state is given by \eqref{therm_lambda}:
\begin{align}\label{2_state_sys}
\begin{split}
	\lambda_0 &= \frac{1}{4} \left[\left( \frac{\coth\frac{\b\w_{0}}{2}}{\w_{0}} + \frac{\coth\frac{\b\w_{1}}{2}}{\w_{1}}   \right)\right. \\
	&~~~~~~~~~~ \left. \times \left(\w_{0}  \coth\frac{\b\w_{0}}{2}   + \w_{1} \coth\frac{\b\w_{1}}{2} \right)  \right]^{1/2}\ ,
\end{split}
\end{align}
where
\begin{align}
\begin{split}
	\w_0 = m, \quad \w_1 = m\sqrt{1+ \frac{4}{m^2\e^2}}\ .
\end{split}
\end{align}
We can plug this into formula \eqref{S_herzog} to get the EE. Instead of giving the general formula, however, we found studying this system in several limits to be illuminating.

\subsubsection{Vacuum and Small Temperature Limit}

When $\b m \to \infty$ for a fixed $m$, the system is in the vacuum, and we have
\begin{align}
\begin{split}
	\lambda_0 &= \frac{1+\sqrt{1+\frac{4}{m^2\e^2}}}{4\left(1 + \frac{4}{m^2\e^2} \right)^{1/4}}\ .
\end{split}
\end{align} 
Notice that $\lambda_0\to 1/2$ for $m\e \to \infty$, which upon substituting into \eqref{S_herzog} implies the EE vanishes. This is the case when the separation distance between neighboring lattice sites diverges.

To make contact with \cite{srednickisub}, we could have expressed our results using $\xi_l$ defined in \eqref{xi_param}. As there's only one symplectic eigenvalue, we have
\begin{equation}
	\xi \equiv \xi_0 = \left( \frac{{\sqrt\w_0} - \sqrt{\w_1}}{{\sqrt\w_0} + \sqrt{\w_1}}  \right)^{2} \ .
 \end{equation}
The EE in this case is given by \eqref{S_xi}, and the reduced subsystem is a single oscillator in a thermal bath with effective temperature $T_{\text{eff}}=\sqrt{\w_0\w_1}/\log(1/\xi)$ \cite{srednickisub}. We verify that the EE we compute indeed matches that obtained in the literature.\footnote{See eq. (6) of \cite{srednickisub} with the identification 
\begin{align}
\begin{split}
	p_i &= \frac{\pi_i}{\sqrt\e}, \quad x_i = \sqrt\e \phi_i, \quad k_0 = m^2, \\
	k_1 &= \frac{2}{\e^2}, \quad \w_{+,-}=\w_{0,1}\ .
\end{split}
\end{align}}

In the small mass limit where $m \ll 1/\e$, the symplectic eigenvalue is 
\begin{equation}\label{zeromode}
\lambda_0=\frac{1}{2{\sqrt{2m\e}}}+\bigO{\sqrt{m\e}}\ ,
\end{equation}
which yields the vacuum EE 
\begin{align}\label{massless_div_2}
	S_A = \frac{1}{2}\log \frac{1}{m\e} + \left(1 - \frac{3}{2}\log 2\right) +{\cal O}(m\e)\ .
\end{align}
The first term is a logarithmic divergence for small mass, while the second one is finite and independent of the mass $m$. 

Next, let us move away from the vacuum and compute the correction to the EE for small temperatures where $T=1/\b \ll m$. Using the asymptotic form of $\coth x=1+2e^{-2x}$ for $x \to \infty$ infinity in \eqref{2_state_sys}, we find 
\begin{equation}
	\delta \lambda_0 =\lambda_0 \left(e^{-\b\w_0}+e^{-\b\w_1}\right)\ , 
\end{equation}
where $\lambda_0$ is the symplectic eigenvalue in the vacuum. Assuming we are far from the decompactification limit, i.e. $m\e \not\to \infty$, the second term is exponentially suppressed and hence can be dropped. We can now use \eqref{var-See} to determine the infinitesimal change in EE away from zero temperature, given that $\delta \lambda_0 \ll \lambda_0 -1/2$. In the small mass limit where $T \ll m \ll 1/\e$, we find
\begin{equation}
	\delta_T S_A\equiv S_A(T) - S_A(0)=\left(1+\bigO{\sqrt{m\e}}\right)e^{-m/T}\ .
\end{equation}
This result is free of divergences, and though we only computed the leading order prefactor, the corrections can easily be computed as well.\footnote{A similar result was also found in \cite{Herzog:2012bw, Cardy:2014jwa} for EEs on line segments.} The result implies that small temperature corrections are exponentially suppressed in the correlators and in the EE, a property that should be true in general.

\subsubsection{Finite Temperature with Small Mass} 

We now turn to the case where we keep the temperature finite and nonzero, and take the small mass limit $m\ll 1/\e $ and $m \ll 1/\b$. Then it follows to leading order
\begin{equation}
	\lambda_0=\frac{1}{2m\beta}\sqrt{1+\frac{\beta}{\e}\coth\frac{\beta}{\e}} + \cdots\ ,
\end{equation} 
with corresponding EE
\begin{align}\label{IRp=2}
\begin{split}
	S_A &=\log\frac{1}{\beta m} + 1 - \log 2 \\
	&~~~~~~~ +\frac{1}{2}\log\left(1+\frac{\beta}{\e}\coth\frac{\beta}{\e}\right) + \cdots \ ,
\end{split}
\end{align}
where $\cdots$ henceforth indicates subleading terms that vanish in the limit we are considering. The first term is divergent at criticality (zero mass limit), and the remaining terms are independent of $m$. 

\subsubsection{High Temperature Limit}

Lastly, let us see what happens in the high temperature limit, where we let $T = 1/\b \gg m$ with $m\e$ fixed. The symplectic eigenvalue to leading order becomes
\begin{align}
\begin{split}
	\lambda_0 &=  \frac{T}{m}\, \sqrt{\frac{ 1+\frac{2}{m^2\e^2}}{1+\frac{4}{m^2\e^2} }} + \cdots\ ,
\end{split}
\end{align}
and the EE is given to be
\begin{align}
\begin{split}
	S_A &= \log\left(\frac{T}{m}\right)+ \frac{1}{2}\log\left(\frac{ 1+\frac{2}{m^2\e^2}}{1+\frac{4}{m^2\e^2} }\right)+1 + \cdots \ .
\end{split}
\end{align}
The massless limit of this expression matches the high temperature limit of \eqref{IRp=2}, as it should. Notice though that while  for small temperatures the EE scales as $e^{-m/T}$, for large temperature the dependence of EE on $m/T$ becomes logarithmic. This is the standard behavior of the Shannon entropy of a Gaussian system in the large temperature limit.

\subsection{Alternating Lattice Entanglement}

Having explored the system of two coupled oscillators, we now generalize to the case when our system consists of $N$ lattice sites with $N = 2N_A$, as depicted on the left in Figure \ref{fig1}. There are now $N_A$ symplectic eigenvalues, with $l = 0,1,\ldots,N_A-1$:
\begin{widetext}
\begin{align}\label{gen_sym_eval}
\begin{split}
	\lambda_l &= \frac{1}{4} \left[\left( \frac{\coth\frac{\b\w_{l}}{2}}{\w_{l}} + \frac{\coth\frac{\b\w_{l+N_A}}{2}}{\w_{l+N_A}}   \right)\left(\w_{l}  \coth\frac{\b\w_{l}}{2}   + \w_{l+N_A} \coth\frac{\b\w_{l+N_A}}{2} \right)  \right]^{1/2}\ ,
\end{split}
\end{align}
\end{widetext}
where $\w_l$ is defined via \eqref{disp} and we used $\w_{l+N_A}=\w_{N_A-l}$ and $\w_{2N_A-l}=\w_{l}$ for the case $p=2$. The EE is as always given via \eqref{S_herzog}, but we again focus our attention on various limits.

\subsubsection{Vacuum and Small Temperature Limit}

\begin{figure}[h]
\centering
{\includegraphics[height=50mm]{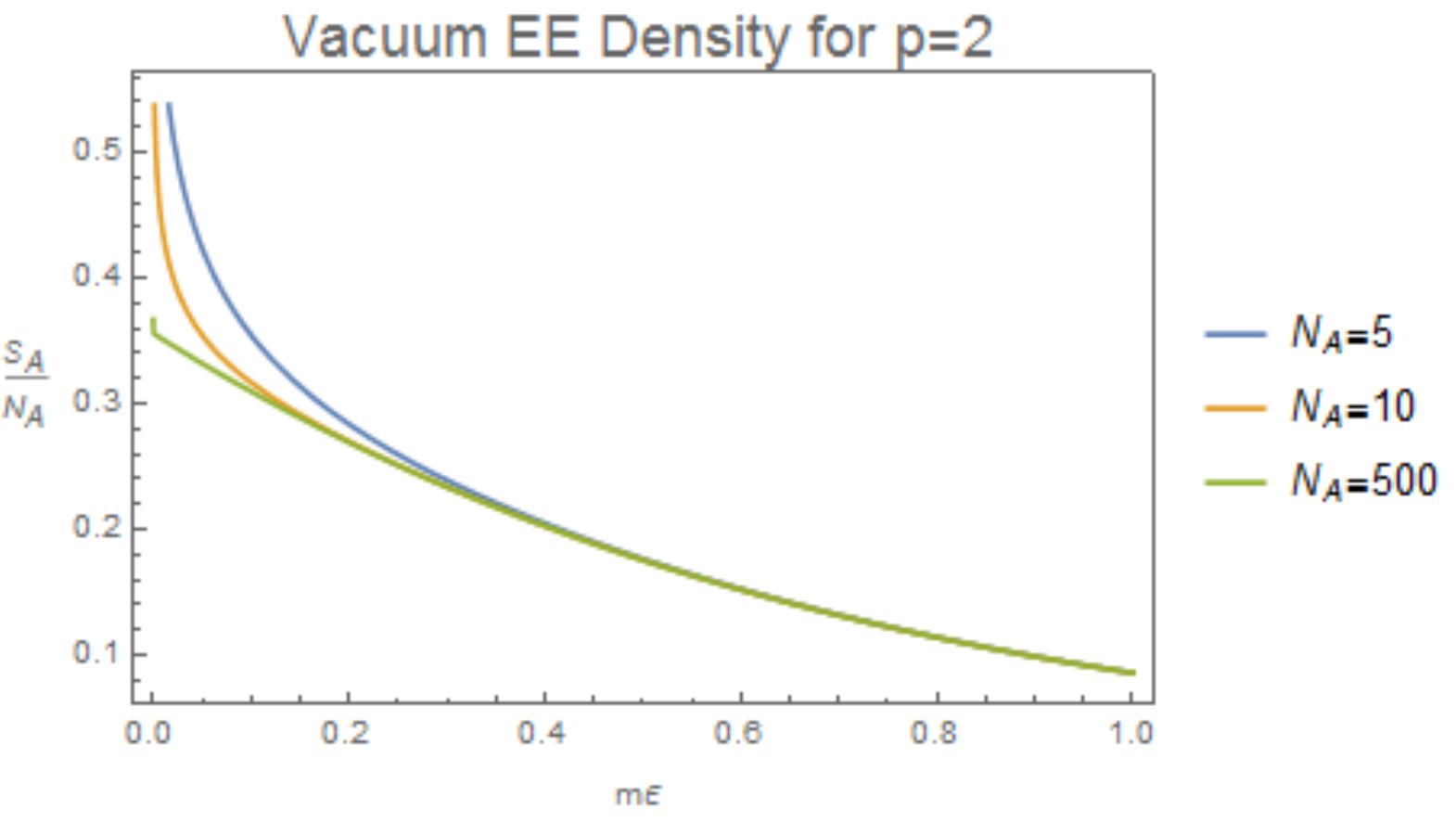}}

\caption{{\small Vacuum EE per lattice site for $p=2$, as a function of $m\e$, for values of $N_A= 5$, 10, and 500. Convergence to extensivity is achieved vary rapidly for sufficiently large $m\e$. The EE is decreasing along the flow $m\e$.}}
\label{figS_ext}
\end{figure}

As before, the limit $\b m\to \infty$ with a fixed $m$ corresponds to the vacuum. The symplectic eigenvalues are
\begin{align}
	\lambda_l = \frac{\w_l+\w_{l+N_A}}{4\sqrt{\w_l\,\w_{l+N_A}}}\ ,
\end{align}
and the vacuum EE by \eqref{S_herzog} is
\begin{align}
\begin{split}
	S_A &= \sum_{l=0}^{N_A-1} \left[-\log\left(4\sqrt{\w_l\w_{l+N_A}} \right) \right. \\
	&~~~~~~~~~\left. + \frac{\left(\sqrt{\w_l} + \sqrt{\w_{l+N_A}} \right)^2}{2\sqrt{\w_l\w_{l+N_A}}}\log\left(\sqrt{\w_l} + \sqrt{\w_{l+N_A}} \right) \right. \\
	&~~~~~~~~~ \left. - \frac{\left(\sqrt{\w_{l+N_A}} - \sqrt{\w_l} \right)^2}{2\sqrt{\w_l\w_{l+N_A}}}\log\left|\sqrt{\w_{l+N_A}} - \sqrt{\w_l} \right| \right]\ .
\end{split}
\end{align}
As the vacuum EE is a function of two variables, $N_A$ and $m\e$, we plotted the EE per lattice site, i.e. the ``EE density,'' versus $m\e$ for various values of $N_A$ in Figure \ref{figS_ext}. We see that the EE is extensive, i.e. the EE density is independent of $N_A$, given sufficiently large $m\e$ and $N_A$. Notice further that the entropy is decreasing as function of the flow parameter $m\e$.

Let us now examine what happens in the limit of small mass, i.e. $m \ll 1/\e$. As in the case where we were only considering two points, $\lambda_0$ diverges again like
\begin{equation}
	\lambda_0=\frac{1}{2}{\frac{1}{\sqrt{2m\e}}}+\cdots
\end{equation}
and contributes to the EE the term
\begin{align}\label{SlogN_A}
	 \frac{1}{2}\log\frac{1}{m\e} + \left(1-\frac{3}{2}\log 2\right) +\cdots\ ,
\end{align}
which yields the same divergent behavior as in the case $N=2$ (see \eqref{massless_div_2}). EE typically diverges when the correlation length is much longer than the lattice spacing $\e$, since this allows for subsystem $A$ to be entangled with many lattice sites in the complement of $A$. As decreasing $m$ translates to increasing the correlation length, small $m\e$ implies the correlation length is much longer than the lattice spacing, and thus as expected we see a divergence in EE coming from the term in \eqref{SlogN_A}.

Thus far, we have ignored the contribution from the other symplectic eigenvalues. Their contribution can be evaluated in the limit where we take the number of points $N$ to be very large but with $N/N_A = 2$ fixed, so we may approximate the sum in \eqref{S_herzog} as an integral. In particular, since $\lambda_l$ is a function of $l/N$, we may denote $\lambda_l \equiv \lambda\left(x \right)$ with $x = l/N$. Excluding the $l=0$ case, which we showed to have a divergence above, we can write
\begin{align}
	\lambda(x) = \frac{\w(x)+\w\left( x + \frac{1}{2}\right)}{4\sqrt{\w(x)\w\left( x + \frac{1}{2}\right)}} =\frac{\sin(\pi x) +\cos(\pi x)}{4{\sqrt{\sin(\pi x)\cos(\pi x)}}}\ ,
\end{align}
where we have taken the limit $m \ll 1/\e$ and kept only the leading term. In the limit of large $N$, $x$ becomes a continuous parameter and the sum over $l\neq 0$ becomes an integral over $x$. The integral is in fact finite, and can be evaluated numerically to be
\begin{align}\label{cts}
\begin{split}
	S_A &= 2N_A\int_{0}^{\frac{1}{2}}\left[\left(\lambda (x)+\frac{1}{2}\right)\log \left(\lambda (x)+\frac{1}{2}\right) \right. \\
	&~~~~~~~~~~~~~~~~~~~ \left. -\left(\lambda (x)-\frac{1}{2}\right)\log \left(\lambda (x)-\frac{1}{2}\right)\right]\,dx \\
	&\simeq 0.36 N_A\ .
\end{split}
\end{align}
We see that the sublattice EE is linear in $N_A$ and hence extensive, as long as the term in \eqref{SlogN_A} is small compared to $N_A$. The EE density is $0.36$ and holds in the large $N$ limit. This corresponds to taking $\e \to 0$ while keeping $L$ fixed and is the UV limit. Referring to Figure \ref{figS_ext}, we see that the EE density indeed approaches this value for large $N_A$ as $m\e \to 0$, although the contribution from $\lambda_0$ still gives a divergence if $m\e$ is sufficiently small, as is also clear in Figure \eqref{figS_ext}. However, the divergence from $\lambda_0$ has a $1/N_A$ suppression when we compute the EE density, so if we take the large $N$ limit faster than $\log(m\e)$, this term does not contribute in the UV. For large values of $m\e$, the correlation length is small compared to the lattice spacing, and hence the EE goes to zero.

Similar to the previous subsection, we can now deviate from the vacuum and increase the temperature infinitesimally. It follows as before
\begin{equation}
	\delta \lambda_l =\lambda_l \left(e^{-\b\w_l}+e^{-\b\w_{l+N_A}}\right)\ . 
\end{equation}
However, it is clear that the dominant term in the sum over $l$ in \eqref{var-See} comes again from the zero mode $\delta \lambda_0$, and the others are exponentially suppressed. Thus, we find in the limit $T\ll m \ll 1/\e$,
\begin{equation}
	\delta_T S_A\equiv S_A(T)-S_A(0)=\left(1+\bigO{\sqrt{m\e}}\right)\,e^{-m/T}\ ,
\end{equation}
which is fully consistent with the case $N_A=1$.

\subsubsection{High Temperature Limit}

Lastly, for the high temperature limit where $\b\w_l \to 0$ for all $l = 0,1,\ldots,N_A-1$, we find 
\begin{align}
\begin{split}
	\lambda_l &= \frac{1}{\sqrt 2 \b}\frac{\sqrt{\w_l^2 + \w_{l+N_A}^2}}{\w_l\w_{l+N_A}}\ .
\end{split}
\end{align}
These diverge for large $1/\b$, so we can use \eqref{Slargelambda} to get 
\begin{align}
\begin{split}
	S_A &= N_A + \sum_{l=0}^{N_A-1} \log\left(\frac{1}{\sqrt 2 \b}\frac{\sqrt{\w_l^2 + \w_{l+N_A}^2}}{\w_l\w_{l+N_A}} \right)\ .
\end{split}
\end{align}
Here we see again a logarithmic growth in $T=1/\b$. For the case where $m$ is large but still smaller than $T$, i.e. $1/\e \ll m \ll T$, all the frequencies are equal and we simply get
\begin{equation}
	S_A=N_A\left(1+\log\left(\frac{T}{m}\right)\right)\ ,
\end{equation}
which is extensive again.

\subsection{$p$-alternating Lattice Entanglement and the Continuum Limit}

\begin{figure}[h]
\centering
{\includegraphics[height=50mm]{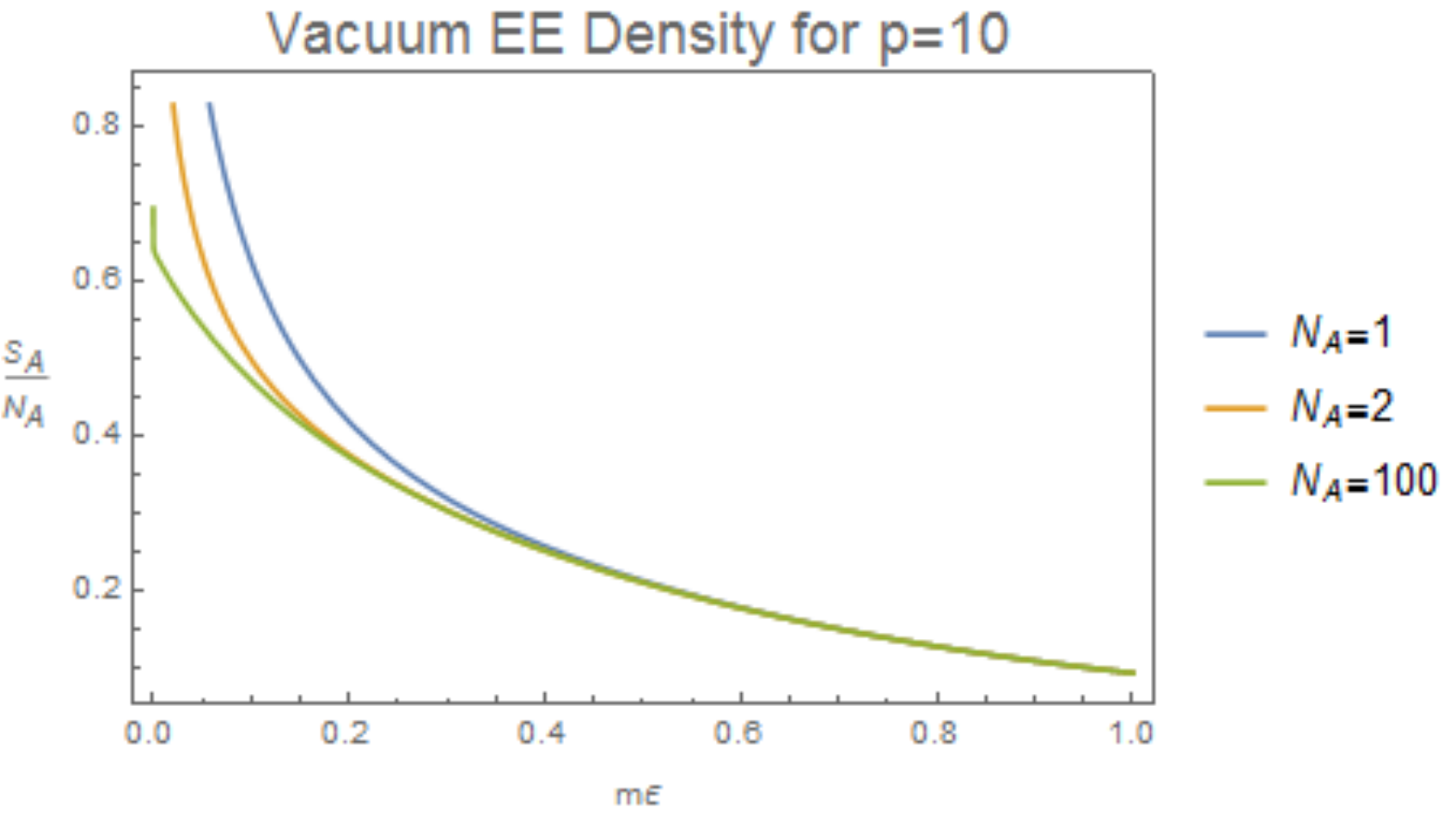}}
\caption{{\small Vacuum EE per lattice site for $p=10$, as a function of $m\e$, for values of $N_A= 1$, 2, and 100 (chosen so that the values for total $N$ is the same as that in Figure \ref{figS_ext}).}}
\label{figS_ext_p=10}
\end{figure}

Finally, we move beyond the $p=2$ case and examine the EE when our subsystem $A$ consists of every $p$-th point on the circular lattice. For any finite $p$, we can straightforwardly extend the results from the previous section. The main difference is that the symplectic eigenvalues are given by \eqref{therm_lambda} with $p>2$ instead of \eqref{gen_sym_eval} and hence more complicated. We can nevertheless make plots, and for $p=10$ the result is given in Figure \ref{figS_ext_p=10}. It is qualitatively similar to the case of $p=2$, but notice that the EE grows faster in the UV limit $m\e\rightarrow 0$.

We can greatly simplify our analysis for the special case of large $p \gg 1$ with $N_A$ fixed, i.e. our subsystem $A$ consists of a very sparse set of points. Studying the EE in this regime is akin to studying how EE behaves in the continuum limit for this particular subsystem of $N_A$ discrete points on the circle. An example of such a system, with $N_A = 4$, is given in Figure \ref{fig2}. 

\subsubsection{Vacuum Limit}

\begin{figure}[h]
\centering
{\includegraphics[height=50mm]{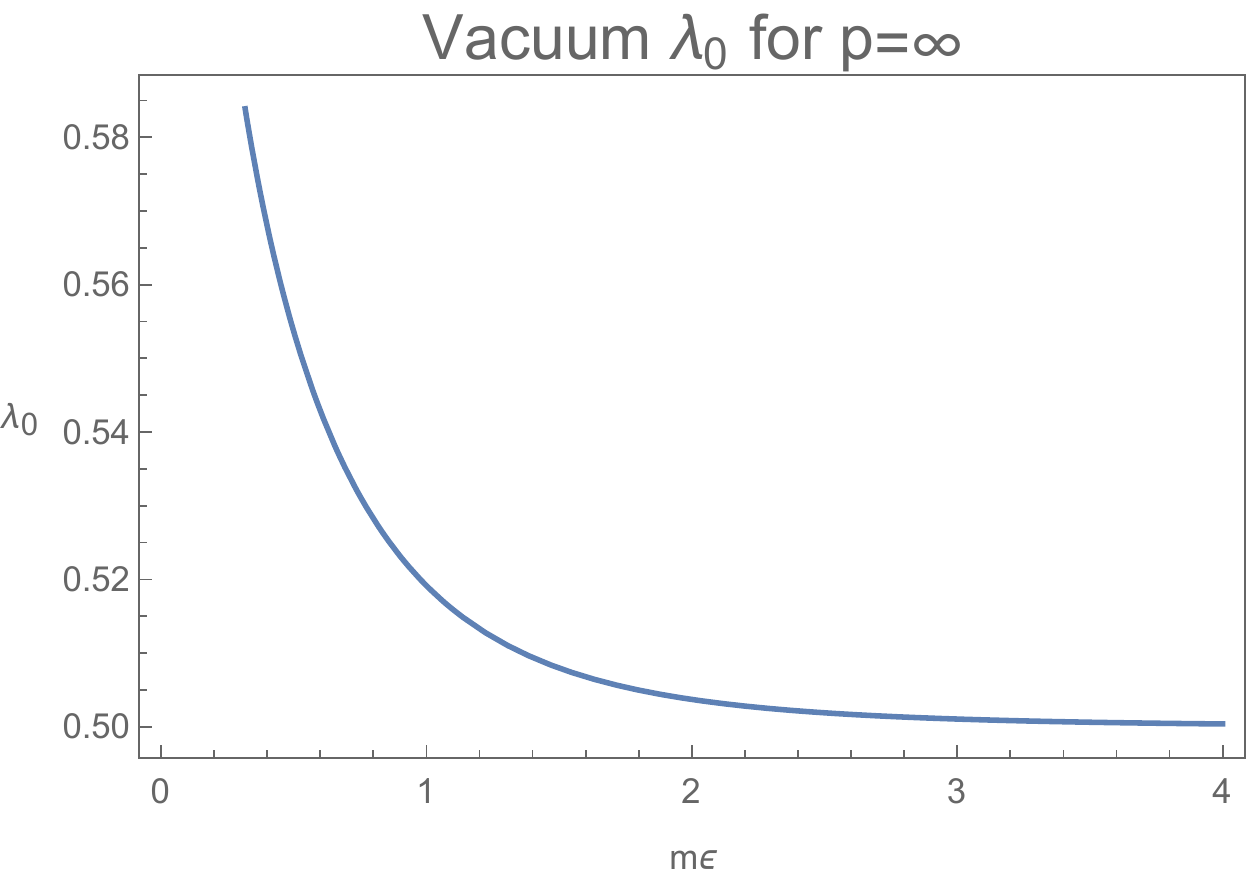}}\hspace{.7cm}
{\includegraphics[height=50mm]{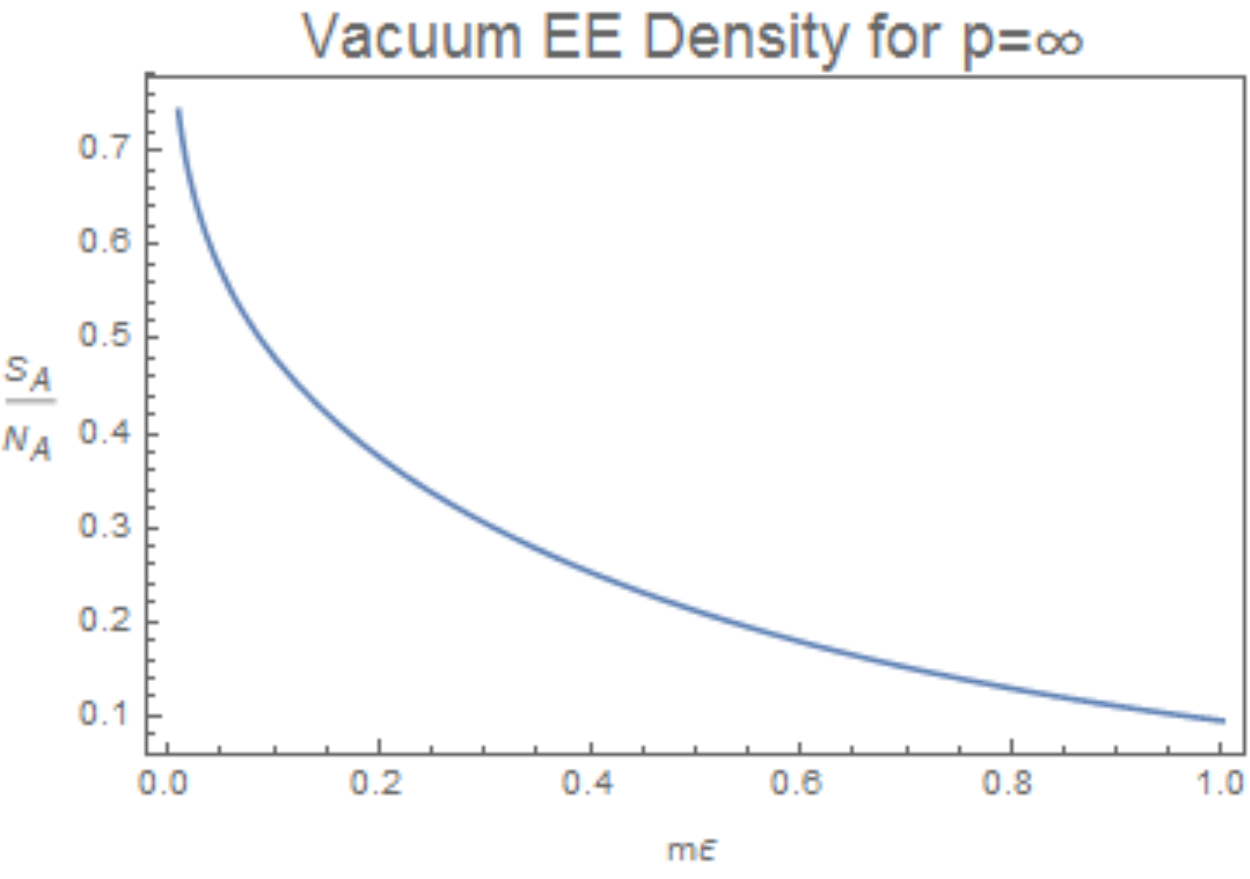}}
\caption{{\small {\em Left}\,: The degenerate eigenvalue $\lambda_0$ in the continuum limit $p, N \rightarrow\infty$. For $m\e \ll 1$ the divergence is slower (logarithmic) than any finite $p$, where it diverges as $(m\e)^{-1/2}$. {\em Right}\,: The corresponding EE for a point on the circle.}}
\label{lambda_0}
\end{figure}

As before, we first discuss the vacuum entanglement, and determine the symplectic eigenvalues using \eqref{mass_lambda} with $p \gg 1$. The crucial observation here is that for large $p$ with fixed $N_A\ll N$, we have $l \ll N$ for all $l$ as $l$ is bounded above by $N_A$. Hence, $l/N \ll1$ and so $\w_{l+kN_A} \approx \w_{kN_A}$, which implies the spectrum of symplectic eigenvalues degenerates, i.e. $\lambda_l \approx \lambda_0$ for all $l$. An immediate consequence of this is that the EE is extensive:
\begin{align}\label{S_pinfty}
\begin{split}
	S_A &= N_A\left[\left(\lambda_0 + \frac{1}{2} \right)\log\left(\lambda_0 + \frac{1}{2} \right) \right. \\
	&~~~~~~~~~~~~~~~~~~~ \left.  - \left(\lambda_0 - \frac{1}{2} \right)\log\left(\lambda_0 - \frac{1}{2} \right) \right]\ ,
\end{split}
\end{align}
and so we only need to determine $\lambda_0$. This can be done because as $N \gg 1$, we may approximate the sums in \eqref{mass_lambda} with integrals from $0$ to $\frac{p}{N} = \frac{1}{N_A}$. Denoting
\begin{align}\label{cont_disp}
	\w(x,y) \equiv \sqrt{m^2 + \frac{4}{\e^2}\sin^2 \pi (x+ N_Ay)}\ ,
\end{align}
we obtain from \eqref{mass_lambda}
\begin{align}\label{lambda_cont}
\begin{split}
	\lambda_0 &= \frac{N_A}{2}\left[\int_0^{1/N_A} \frac{1}{\w(0,y)}\,dy\,  \int_0^{1/N_A}  \w(0,y) \,dy \right]^{1/2} \\
	&= \frac{1}{\pi}\sqrt{E\left( - \frac{4}{m^2\e^2} \right)K\left( - \frac{4}{m^2\e^2}\right)}\ ,
\end{split}
\end{align}
where $E$ and $K$ are the standard complete elliptic integrals. A plot of $\lambda_0$ as a function of $m\e$, as well as the corresponding EE for a point, is given in Figure \ref{lambda_0}. Notice that for $m\e \gg 1$, $\lambda_0$ approaches the minimum value 1/2, and the contribution from such symplectic eigenvalues to the EE vanishes.

To study the behavior of the EE near $m\e \ll 1$, we Taylor expand \eqref{lambda_cont} around small $m\e$ to obtain
\begin{align}
\begin{split}
	\lambda_0 &= \frac{1}{\pi}\sqrt{\log 8 + \log\frac{1}{m\e}} + \cdots\ .
\end{split}
\end{align}
This has a logarithmic divergence for $m\e\ll 1$, and the corresponding EE using \eqref{Slargelambda} is
\begin{align}\label{S_lowT}
\begin{split}
	S_A =  N_A \left[ \frac{1}{2}\log\left(\log\frac{1}{m\e}\right) + 1 - \log \pi +  \cdots \right]\ .
\end{split}
\end{align}
As we mentioned earlier, because $m\e$ measures the ratio of the lattice spacing to the correlation length, a small $m\e$ naturally implies more entanglement. Notice though that for this case, the divergence is a much softer double logarithm, whereas for finite $p$, the divergence is logarithmic. 

\subsubsection{High Temperature Limit}

We can perform a similar analysis in the high temperature limit. As in the vacuum case, all the symplectic eigenvalues degenerate in the continuum limit and we obtain from \eqref{therm_lambda} 
\begin{align}
\begin{split}
	\lambda_0 &= \frac{N_A}{2}\left[\int_0^{1/N_A} \frac{\coth\frac{\b\w(0,y)}{2}}{\w(0,y)} \,dy \right. \\
	&~~~~~~~~~~~~~~~  \left. \times \int_0^{1/N_A} \w(0,y)\coth\frac{\b\w(0,y)}{2} \,dy \right]^{1/2}\ .
\end{split}
\end{align}
Taking the high temperature limit $\b \w(0,y) \to 0$ for all $y \in [0,1/N_A]$, we obtain
\begin{align}
\begin{split}
	\lambda_0 = \frac{1}{\b m\left(1+\frac{4}{m^2\e^2}\right)^{1/4}} + \cdots \ .
\end{split}
\end{align}
In the high temperature regime, $\b m \to 0$, so $\lambda_0$ diverges. Using \eqref{S_pinfty} for large $\lambda_0$, we obtain to leading order in $m\e$ and $\beta m$,
\begin{align}\label{S_highT}
\begin{split}
	S_A =  N_A \left[\log\frac{T}{ m} - \log\frac{1}{m\e} + 1 - \frac{1}{2}\log 2 +\cdots \right]\ .
\end{split}
\end{align}

\subsection{Mutual Information}

In this subsection, we briefly comment on one interesting quantity we may compute: the mutual information between two opposite points on the circle. While it is possible to work this out for arbitrary temperatures using the formulas we derived above, we restrict ourselves to the vacuum case where the expressions are simple. The extension to nonzero temperatures should be straightforward.

Let us start with $N$ lattice sites on the circle where $N$ is even. The mutual information between two subsystems $A$ and $B$ is by definition
\begin{align}\label{Iab}
	I_{AB} = S_A + S_B - S_{AB}\ ,
\end{align}
where $AB$ denotes the union $A \cup B$. Mutual information is free of divergences and positive. For our interests, we take $A$ to be a single lattice site and $B$ to be the opposite lattice site on the circle. As $S_A = S_B$, let us focus on subsystem $A$, i.e. the case where $N_A = 1$ and $p=N$. Applying \eqref{mass_lambda}, we note that there is only one symplectic eigenvalue, namely
\begin{align}
	\lambda_0^A = \frac{1}{2N}\sqrt{\sum_{j,k=0}^{N-1}\frac{\w_j}{\w_k}}\ .
\end{align}
Substituting this directly into \eqref{S_herzog} yields $S_A$.

\begin{figure}[h]
\centering
{\includegraphics[height=50mm]{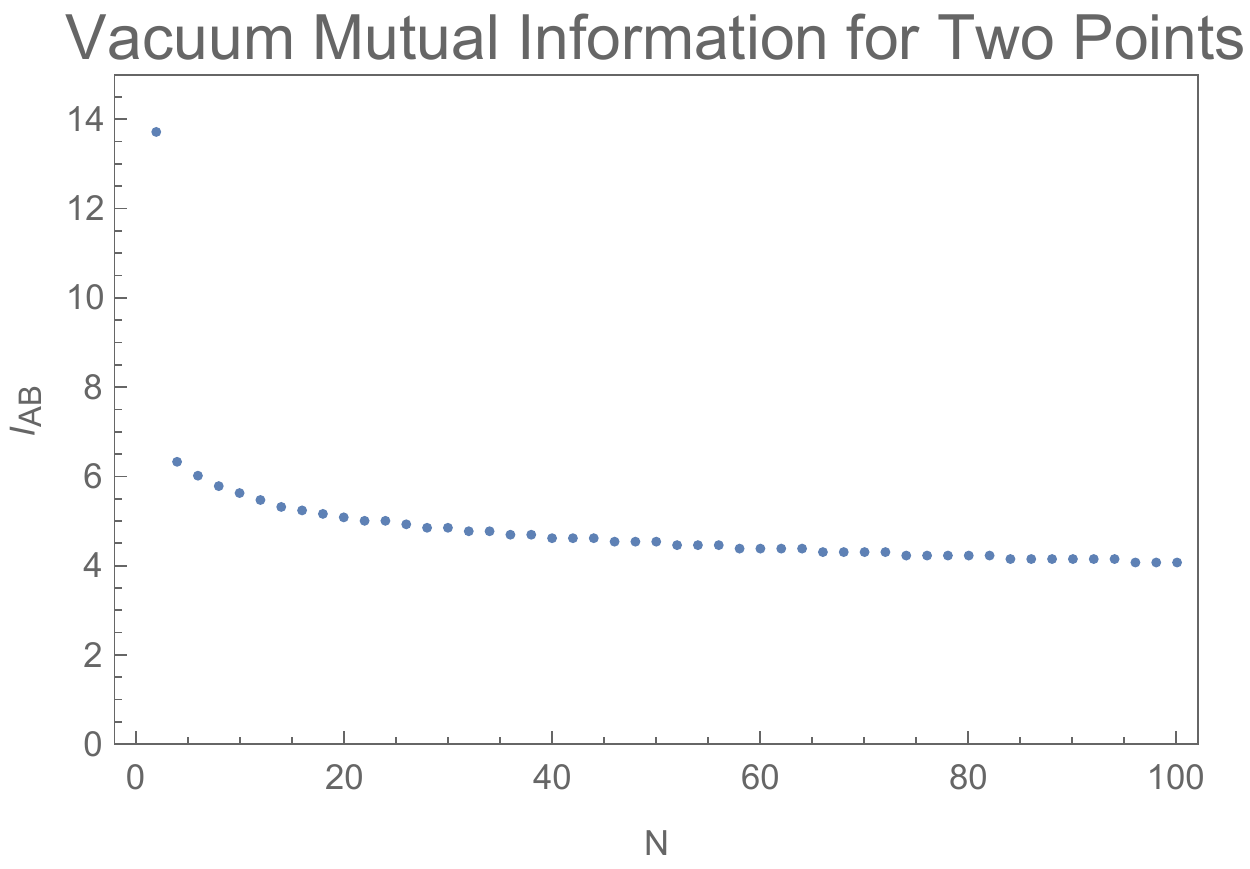}}

\caption{{\small A plot of the mutual information between two opposite points in the circular lattice with $N$ lattice sites with $m\e = 10^{-6}$ and $T = 1/\b = 0$. In the limit $N \to \infty$ the mutual information goes to zero, albeit rather slowly.}}
\label{mutual_info}
\end{figure}

Similarly, we can compute the EE of $S_{AB}$. This corresponds to the case $N_A = 2$, so $p=N/2$ and there are now two symplectic eigenvalues:
\begin{align}
\begin{split}
	\lambda_0^{AB} &= \frac{1}{N}\sqrt{\sum_{j,k=0}^{\frac{N}{2}-1} \frac{\w_{2j}}{\w_{2k}}}\ , \qquad  \lambda_1^{AB} = \frac{1}{N}\sqrt{\sum_{j,k=0}^{\frac{N}{2}-1} \frac{\w_{2j+1}}{\w_{2k+1}}  }\ .
\end{split}
\end{align}
Again, direct substitution into \eqref{S_herzog} yields $S_{AB}$. We can now for any finite even $N$ obtain the mutual information via \eqref{Iab}, and we plotted the result in Figure \ref{mutual_info} for the case where $m\e$ is fixed. 
Notice that $I_{AB}$ decreases monotonically as the number of lattice sites between $A$ and $B$ increases, as we expect. Indeed, in the continuum, the EEs $S_A$ and $S_B$ are simply given by \eqref{S_lowT} for $N_A = 1$, and $S_{AB}$ given for $N_A = 2$. It immediately follows that $I_{AB} = 0$, i.e. the infinitely many lattice sites between $A$ and $B$ have completely decoupled $A$ from $B$. Further numerical analysis for very large values of $N\sim 10^6$ shows that this is indeed the case.

\section{Discussion}\label{discussion}

For Gaussian systems, two-point correlation matrices that are circulant on a periodic sublattice lead to entanglement spectra and EEs that can be computed exactly. This has been explicitly illustrated in this paper for a massive scalar field theory as a particular example. Various extensions and generalizations are possible. Firstly, an obvious extension is to apply the methods to fermionic systems such as spin chains. Some models were already studied in \cite{IP}, but a general analysis of EEs of Gaussian spin/fermionic systems at arbitrary temperatures is to our knowledge not known. Secondly, the matrix of couplings $V$ in the potential terms of the Hamiltonian \eqref{discrete-Ham}, written as $\phi_iV_{ij}\phi_j$ is of the nearest-neighbor type. One can extend this to any circulant matrix to include for instance long-range interactions. Then one can compute how the EE changes in the presence of these long range interactions. One intriguing case, for instance, is to take all matrix entries in $V$ equal, so that all the lattice sites are equally connected. Finally, one can try to extend our formalism to higher dimensions. For instance, in two spatial dimensions, we can take a finite square lattice with periodic boundary conditions, i.e. a discretized torus. We can then take as a periodic sublattice every $p$-th point in both directions, as long as $p$ divides the number of lattice sites in both directions. It would be interesting to work out explicit examples of these extensions and unravel the dependence on the background geometry of this notion of entropy.

Throughout this paper, the question of what universal information is actually contained inside this $p$-alternating entropy quantity has not been discussed. It is obvious that the computed entanglement spectrum given by \eqref{ex_sym} depends on all the parameters of the microscopic theory, so it is possible that this quantity contains the universal information pertaining to a phase transition (which is obviously contained within the microscopic theory). However, extracting such universal information from the $p$-alternating entropy might be more difficult than from the usual notion of subsegment entropy, where the subsystem consists of contiguous lattice sites. This is because it is not obvious how to isolate the regulator dependence for the $p$-alternating entropy, and we leave this as an interesting open question that will hopefully be further explored in the future.

We stress that the framework presented here can be extended to any system with a Gaussian correlator structure. One might think that such feature constrains the sytem to be free of interactions. As discussed in \cite{usdemo}, this is not the only case, the crucial and paradigmatic example being large-$N$ gauge theories. A characteristic feature of such models is large-$N$ factorization \cite{largeN}, which implies an effective emergent Gaussianity for reduced subsystems, even when the system is strongly interacting. In \cite{usdemo}, such large-$N$ factorization was exploited in Fock space, where it implies extensivity of entanglement evolution (see \cite{pawel} for another study of entanglement dynamics in non-local systems). It would then be interesting to apply the methods developed in this paper to the case of large-$N$ matrix models. A paralell approach to such effective gaussianity appears when using random unitaries, as in \cite{vijay,MV}. This approach is simpler and it is a good starting point for the more complicated matrix models.

\section*{Acknowledgements}

It is a pleasure to thank Y.-H. Lin, P. Mitra, A. Queiroz, S. Paganelli, D. Schuricht, B. Schwab, and A. Strominger for interesting discussions.
This work was supported by the Netherlands Organisation for Scientific Research (NWO) under the VICI grant 680-47-603, and the Delta-Institute for Theoretical Physics (D-ITP) that is funded by the Dutch Ministry of Education, Culture and Science (OCW).





\end{document}